\documentclass[english]{article}
\usepackage[latin9]{inputenc}
\usepackage{color}
\usepackage{textcomp}
\usepackage{amstext}
\usepackage{amssymb}
\usepackage{esint}

\makeatletter

\DeclareRobustCommand{\greektext}{%
  \fontencoding{LGR}\selectfont\def\encodingdefault{LGR}}
\DeclareRobustCommand{\textgreek}[1]{\leavevmode{\greektext #1}}
\DeclareFontEncoding{LGR}{}{}
\DeclareTextSymbol{\~}{LGR}{126}
\newcommand{\lyxmathsym}[1]{\ifmmode\begingroup\def\b@ld{bold}
  \text{\ifx\math@version\b@ld\bfseries\fi#1}\endgroup\else#1\fi}

\newcommand{\lyxaddress}[1]{
\par {\raggedright #1
\vspace{1.4em}
\noindent\par}
}

\makeatother

\usepackage{babel}
\begin{document}

\title{\textbf{Noncommutative geometry of AdS coordinates on a D-brane}}

\author{\textbf{Lawrence B. Crowell}}

\maketitle

\lyxaddress{Alpha Institute of Advanced Study 10600 Cibola Lp 311 NW Albuquerque,
NM 87114 also 11 Rutafa Street, H-1165 Budapest, Hungary, email: \textcolor{blue}{lcrowell@swcp.com}; }

In this short paper the noncommutative geometry and quantization of
branes and the AdS is discussed. The question in part addresses an
open problem left by this author in {[}1{]} on how branes are generated
by stringy physics. The breaking of an open type I string into two
strings generates a nascent brane at the new endpoints with inflationary
cosmologies. This was left as a conjecture at the end of this paper
on the role of quantum critical points in the onset of inflationary
cosmology. The noncommutative geometry of the clock and lapse functions
for the AdS-brane are derived as is the number of degrees of freedom
which appear. The role of the AdS spacetime, or in particular its
boundary, in cosmology is discussed in an elementary regularization
scheme of the cosmological constant on the boundary. This is compared
to schemes of conformal compactification of the AdS spacetime and
the Heisenberg group.

\section{Geometry of QCD theory of D-brane with AdS coordinates}

In the recent article {[}1{]} a quantum phase transition model for
the onset of inflation is proposed. The quantum critical behavior
is proposed as a change in the physics of a type I string attached
to two D-branes. As the D-branes separate under Casimir vacuum pressure
the string stretches and breaks, with a new D-brane holding the endpoints
of the two strings. The emergence of a D-brane is a quantum to classical
transition. D-branes are classical objects which emerge in the limit
of large ${\cal N}$ modes or degrees of freedom on the brane. In
this letter the physics of this phase transition is examined in the
light of holographic bounds.

The open string is a near Planck scale version of a meson. The endpoints
are quark-like particles with an analogue of a gluon flux tube connecting
them, which serves as the string. The system is a type of two-quark
QCD system. The endpoints or quarks exist in a family of $N_{f}$
quark fields and we represent this theory as $SU(2N_{f})_{r}\times SU(2N_{f})_{\ell}$.
The general Lagrangian for a QCD system such as this {[}2{]} is

\[
{\cal L}~=~-\frac{1}{4\pi g^{2}}F_{\mu\nu}^{a}F^{a\mu\nu}~+~i\bar{\psi}\sigma^{\mu}\Big(\partial_{\nu}~+~\frac{i}{2}A_{\mu}^{a}\tau^{\mu}\Big)\psi~-~\frac{1}{2}m_{q}\psi^{T}\tau_{2}{\bf \Omega}\psi~+~HC
\]
for $\psi$ the two component spinor 

\[
\psi~=~\left(\begin{array}{c}
q_{\ell}\\
\sigma_{2}\tau_{2}q_{r}
\end{array}\right)
\]
and $A_{\mu}^{a}$ the gluon field strength with chromo-index $a~=~1,~2,~3$.
The matrix ${\bf \Omega}$ is the $2N\times2N$ skew symmetric matrix 

\[
{\bf \Omega}~=~\left(\begin{array}{cc}
{\bf 0} & {\bf 1}\\
-{\bf 1} & {\bf 0}
\end{array}\right)
\]
The massless limit with $m_{q}~=~0$ the $SU(2N_{f})$ symmetry is
replaced with $U(2N_{f})$ for $Sp(n)~=~U(2n)\cap Sp(2n,~\mathbb{C})$. 

The Hermitian generator of $SU(2N_{f})$ of dimension $2N_{f}~-~1$
are normalized as $Tr(T^{a}T^{b})~=~\delta^{ab}/2$ exist in two sets.
The first set pertain to the symplectic group $Sp(2N_{f})~\subset~SU(2N_{f})$,
denoted as $X^{a}$, $a~=~1,~\dots,~2N_{f}~-~N_{f}$, and the remainder
$Y^{a}$ pertain to the quotient group $SU(2N_{f})/Sp(2N_{f})$ for
$a~=~1,~\dots,~2N_{f}~+~N_{f}~-~1$. The quotient group generators
{}``left over'' from the group reduction are the Goldstone bosons
in the $2N_{f}\times2N_{f}$ matrix

\[
{\bf Z}~=~e^{in_{a}x^{a}/\sqrt{N}}{\bf \Omega}
\]
The algebraic elements of $Sp(2N_{f})$ group obey

\[
{\bf X}^{T}{\bf \Omega}~+~{\bf \Omega X}~=~0
\]
and the quotient group obey

\[
{\bf Y}^{T}{\bf \Omega}~-~{\bf \Omega Y}~=~0.
\]

An important example is the group $SU(4)$, since $SU(4)~\sim~SO(4,~2)$
is the isometry group of $AdS_{4}~\sim~SO(4,~2)/SO(4,~1)$. The generators
$X^{a}$ and $Y^{a}$ of $Sp(2,~2)~\sim~Sp(4)$ and the quotient subgroup
in $SU(4)$ can be written as 

\[
X^{a}~=~\frac{1}{2\sqrt{2}}\left(\begin{array}{cc}
\sigma^{a} & 0\\
0 & -\sigma^{aT}
\end{array}\right){}_{a=1,\dots4},~X^{a}~=~\frac{1}{2\sqrt{2}}\left(\begin{array}{cc}
0 & x^{a}\\
x^{a\dagger} & 0
\end{array}\right){}_{a=5,\dots10},~
\]

\[
Y^{a}~=~\frac{1}{2\sqrt{2}}\left(\begin{array}{cc}
\sigma^{a} & 0\\
0 & \sigma^{aT}
\end{array}\right)_{a=1,\dots3}~Y^{a}~=~\frac{1}{2\sqrt{2}}\left(\begin{array}{cc}
0 & y^{a}\\
-y^{a\dagger} & 0
\end{array}\right)_{a=4,5}
\]
$X^{a}$ can be seen from the $Sp(4)$ identity ${\bf X}^{T}{\bf \Omega}~+~\Omega{\bf X}~=~0$.
$\sigma^{a}$ are the standard Pauli matrices for $a~<~4$, and for
$a~=~4$ this is a unit matrix. The group is then $SU(2)\times U(1)$.
For $a~=~5,~\dots,~10$ the elements are

\[
x^{5}~=~1,~x^{7}~=~\sigma^{3},~x^{9}~=~\sigma^{1},~x^{a+1}~=~ix^{a}
\]
The $Y^{a}$ elements are seen from for ${\bf Y}^{T}{\bf \Omega}~-~\Omega{\bf Y}~=~0$
and the elements $y^{4}~=~\sigma^{2}$, $y^{5}~=~i\sigma^{2}$.

Now decompose the matrix ${\bf Z~=~{\bf U~+~{\bf V}}}$ with

\[
U^{a}~=~\frac{1}{2\sqrt{2}}e^{-2\sqrt{2}}\left(\begin{array}{cc}
0 & e^{\sigma^{a}/\sqrt{n}}\\
0 & 0
\end{array}\right),~V^{a}~=~\frac{1}{2\sqrt{2}}e^{-2\sqrt{2}}\left(\begin{array}{cc}
0 & 0\\
-e^{-\sigma^{aT}/\sqrt{n}} & 0
\end{array}\right)
\]
with the result that the multiplication of the two matrices is 

\[
U^{a}V^{b}~=~\frac{1}{8}e^{-2\sqrt{2}}\left(\begin{array}{cc}
-e^{\sigma^{a}/\sqrt{n}}e^{-\sigma^{bT}/\sqrt{n}} & 0\\
0 & 0
\end{array}\right),
\]

\[
V^{b}U^{a}~=~\frac{1}{8}e^{-2\sqrt{2}}\left(\begin{array}{cc}
0 & 0\\
0 & -e^{-\sigma^{bT}/\sqrt{n}}e^{\sigma^{a}/\sqrt{n}}
\end{array}\right)
\]

The product $e^{\sigma^{a}/\sqrt{n}}e^{-\sigma^{bT}/\sqrt{n}}~\simeq$
$e^{(\sigma^{a}~-~\sigma^{bT})/\sqrt{n}}e^{[\sigma^{a},~\sigma^{bT}]/2n}$
and the commutator is 

\[
U^{a}V^{b}~-~V^{b}U^{a}~=~\frac{1}{8}e^{-4\sqrt{2}}e^{(\sigma^{a}~-~\sigma^{bT})/\sqrt{n}}\left(\begin{array}{cc}
-e^{[\sigma^{a},~\sigma^{bT}]/2n} & 0\\
0 & e^{-[\sigma^{a},~\sigma^{bT}]/2n}
\end{array}\right)
\]
The transpose of the Pauli matrix $\sigma^{2T}~=~-\sigma^{2}$ with
the rest remaining the same means the commutator in the matrix is$[\sigma^{a},~\sigma^{bT}]~=~2i\epsilon^{abc}\sigma^{cT}$
and we have for $a~=~+$ and $b~=~-$ that

\[
U^{+}V^{-}~-~V^{-}U^{+}~=~\frac{1}{8}e^{-4\sqrt{2}}e^{(\sigma^{+}~-~\sigma^{-T})/\sqrt{n}}\left(\begin{array}{cc}
-e^{i\sigma^{z}/n} & 0\\
0 & e^{-i\sigma^{z}/n}
\end{array}\right)
\]
or approximately

\[
U^{+}V^{-}e^{-i\sigma^{z}/n}~-~V^{-}U^{+}e^{i\sigma^{z}/n}~=~\frac{1}{8}e^{-4\sqrt{2}}e^{(\sigma^{+}~-~\sigma^{-T})/\sqrt{n}}\left(\begin{array}{cc}
-1 & 0\\
0 & 1
\end{array}\right)
\]
which leads to

\[
U^{+}V^{-}e^{-i\sigma^{z}/n}~-~V^{-}U^{+}e^{i\sigma^{z}/n}~=~0
\]
If we reset $1/n~\rightarrow~\pi/n$ and evaluate this matrix on the
eigenvector $|+\rangle$ of $\sigma^{z}$ we can write this result
as $U^{+}V^{-}~-~V^{-}U^{+}e^{2\pi i/n}~=~0$. This construction of
noncommutative geometry is a $\star$-product extension of a symplectic
geometry.

\section{Clock-shift operators under Lorentz boost and degrees of freedom}

This means the manifold is a {}``fuzzy'' space with noncommutative
geometry. The operators $U^{\pm},~V^{\pm}$ are spinor versions of
the clock and shift functions on an $n~=~2$ dimensional Hilbert space
{[}3{]}. The structure is then a reduced version of the large N version
of the noncommutative coordinates of a D-brane. This theory may be
extended into a Witt algebra, or a Virasoro algebra. The Pauli matrices
are elements of $SU(2)~\sim~SO(3)$. The Lorentzian form of this theory
is $SU(1,~1)~\sim~SO(2,~1)$. The projective $2~+~1$ Lorentz group
$PSL(2,~1)$ is isomorphic to the $SL(2,~\mathbb{R})$ defined for
the operators $L_{1},~L_{-1}$ and $L_{0}$ by

\[
[L_{0},~L_{-1}]~=~L_{-1},~[L_{0},~L_{1}]~=~-L_{1},~[L_{1},~L_{-1}]~=~2L_{0}.
\]
where these operators are expanded in modes as according to the Laurent
expansion 

\[
L^{n}~=~\oint\frac{dz}{2\pi iz}z^{n+2}T(z),~T(z)~=~-\sum_{n=\infty}^{\infty}\frac{L^{n}}{z^{n+2}}.
\]
The $SL(2,~\mathbb{R})$ algebra may be embedded into a Virasoro algebra
{[}4{]}

\[
[L_{m},~L_{n}]~=~(m~-~n)L^{m+n}~+~c(m)\delta_{ij}.
\]
The anomaly term is $c(m)~=~D(m^{3}~-~m)/12$, for $D~=~26$. The
states of the system are then given by $L_{m}~=~\sum_{n}\alpha_{m-n}\alpha_{m}$,
and if the group is restricted to $SL(2,~\mathbb{R})$ the mode operators
which form this algebra sum accordingly. The $L_{0}$ portion of the
$SL(2,~\mathbb{R})$ is the operator $L_{0}~=~\sum_{n}\alpha_{-n}\alpha_{n}$
which is the Hamiltonian for the bosonic string. 

We have extended this construction to a larger Hilbert space, where
if $c(n)~=~0$ for all $n$ this is the Witt algebra for $L_{n}~=~-z^{n+1}\partial/\partial z$.
For the Witt algebra over a finite field the largest $L_{N}$ value
would correspond to the upper frequency limit on the Hilbert space
of $N$ dimensions. the finite Witt algebra is some imposed by a time
resolution in the observation of the D-brane. The D-brane is composed
of cells of minimal uncertainty with $[p,~x]~=~\hbar$. The uncertainty
in the momentum is given by a resolution time $\delta t~=~\epsilon$.
We may then remove any energy-momentum greater than $1/\epsilon$.
On the infinite momentum frame the energy is $E~=~(p_{\perp}^{2}+~m^{2})/2P$,
for $P$ the longitudinal momentum. This conversely means the longitudinal
momentum must be $P~<~m^{2}\epsilon$. A D-brane with $N$ degrees
of freedom is then determined by the longitudinal boost of that brane
relative to another brane. A degrees of freedom on the brane are increased
by Lorentz boosting the brane, such as doubling the momentum means
$P~<~2m^{2}\epsilon$, and the number of energy states on the brane
has increased. The boost in the brane increases the resolution time
by the dilation of time so a new set of degrees of freedom appear
in the energy region $m^{2}\epsilon~<~P~<~2m^{2}\epsilon$. The Witt
algebra over a finite field is then extended. The Witt algebra over
a field $k[z]$ of characteristic $N~>~0$, is the Lie algebra of
derivations of the ring $k[z]/z^{N+2}$ The Witt algebra is spanned
by $L_{n}$ for $\lyxmathsym{\textminus}1~\le~n~\le~N$. The boost
$P~\rightarrow~2P$ redefines the ring to $k[z]/z^{2N+2}$, and the
increase in the number of modes or degrees of freedom is a manifestation
of the Lorentz boost factor. This is a form of generating Feynman's
wee-partons {[}5{]}.

A QCD-like string with the field $A_{\mu}^{a}$ interacting on branes
with a QCD-like chromocharge and open ends as fermion fields $\psi$,
or quarks on these branes, is broken as the two branes separate. The
separating endpoints are connected to a nascent brane with a few degrees
of freedom. This nascent brane is a $S^{3}$ corresponding to a FLRW
metric, which expands to its turn around or maximum expansion point.
The violation of the Bekenstein bound at the turn around point forces
this surface to becomes $\mathbb{R}^{3}$. Equivalently the $S^{3}$
becomes enormously Lorentz boosted relative to the end points and
their Dirichlet boundary conditions on the brane and on the infinite
momentum frame appears as a stretched horizon. The metric on this
surface is an anti de Sitter spacetime. The transverse modes of the
string become enormously Lorentz boosted relative to the nascent brane,
which is a form of stretched horizon as measured by an observer near
either original endpoints of the string. The transverse modes of the
string increase and the string covers the nascent brane increasing
the number of modes observed on it. The appearance of the stretched
horizon covered by the string means each region of the surface with
a Planck unit of area $G\hbar/c^{3}$ contain a mode in the limit
$N~\rightarrow~\infty$. The two operators $U$ and $V$ are then
elements of an enveloping algebra of complementary observables with
a minimal uncertainty $\hbar$. 

With the construction with Pauli matrices we have the $SU(2)$ commutation
relationship for angular momentum $[L_{i},~L_{j}]~=~i\epsilon\hbar L_{k}$.
Now choose a coordinate system on the sphere with $L_{z}$ through
the origin. Then $L_{z}~\simeq$\textasciitilde{} constant and we
may write $L_{x}~=~px$ and $L_{y}~=~py$ so that $[x,~y]~=~i\hbar\theta$.
The same applies for the momentum space. The conversion to this construction
with $\sigma^{2T}$ then maps this sphere into the hyperbolic coordinates
considered. This gives a meaning to the noncommutativity of the coordinates
of the manifold. 

\[
e^{i\nabla_{i}}e^{i\nabla_{j}}=e^{i\nabla_{i}+i\nabla_{j}+\frac{1}{2}R_{ijkl}y^{i}y^{k}}.
\]
The Riemann curvature pertains to the AdS Riemann curvature tensor
components. In addition the curvature here is in $O(\hbar/N)$ and
is then a quantized effect. The world volume swept out by the D3-brane
is defined by the $AdS_{4}$ curvatures in $t,~\chi,\theta,~\phi$
coordinates

\[
R_{t\chi t\chi}~=~cos^{2}(t),~R_{t\theta t\theta}~=~cos^{2}(t)sinh^{2}(\chi),~R_{t\phi t\phi}~=~cos^{2}(t)sinh^{2}(\chi)sin^{2}(\theta)
\]

\[
R_{\chi\theta\chi\theta}~=~-cos^{2}(t)sinh^{2}(\chi)~+~cos(t)sin(t)sinh^{2}(\chi)
\]

\[
R_{\chi\phi\chi\phi}~=~-cos^{2}(t)sinh^{2}(\chi)sin^{2}(\theta)~+~cos^{2}(t)sin^{2}(t)sinh^{2}(\chi)sin^{2}(\theta)
\]

\[
R_{\theta\phi\theta\phi}~=~cos^{2}(t)sinh^{2}(\chi)sin^{2}(\theta)~+~cos^{2}(t)sin^{2}(t)sinh^{4}(\chi)sin^{2}(\theta)
\]

\[
~-~cos^{2}(t)sinh^{2}(\chi)sin^{2}(\theta)cosh^{2}(\chi)
\]
The last three of these curvatures are the curvature of the spacetime
on the D3-brane, while the first three above are the curvature of
the D3-brane in the world volume it sweeps through. 

In this setting the $U$ and $V$ operators with a commutation given
by $UV~=$ $e^{2\lyxmathsym{\textgreek{p}}i/N}VU$ and the deviation
from commutation is determined by this curvature in units of $\hbar/N$.
Equivalently we may think of the variables $\chi~\rightarrow~\chi/\sqrt{N}$,
where in the limit $N~\rightarrow~\infty$ the curvatures approach
zero. The number of degrees of freedom in the system is {}``large
$N$,'' not infinite. Consequently the anti de Sitter spacetime on
the D-brane {}``matures'' into a state with curvature present only
over considerable distances on the brane. A realization of clock and
shift operators is a noncommutative geometry on the brane, and large
$N$ corresponds to a high boost of the brane and a classical limit.

\section{Is the observable cosmology anti de Sitter, or the boundary of $AdS_{n}$?}

Hartle, Hawking and Hertog {[}5{]} have suggested the observable universe
may indeed by anti de Sitter. With the Wheeler DeWitt equation they
derive an expanding wave function in an AdS spacetime with the energy
constraint

\[
\left(\frac{a'}{N}\right)^{2}~-~1~-~\frac{a^{2}}{\ell^{2}}=0,
\]
where the sign change is such that $\ell^{-2}~=~-\Lambda/3$ for the
AdS. The negative cosmological constant makes the relationship between
quantum physics and gravity far easier to understand as the AdS/CFT
correspondence and holography. The quantum wave functional of the
Wheeler-DeWitt equation expands with the scale factor $a$, which
it is argued would correspond to an expanding universe. 

The $AdS_{4}$ metric

\[
ds^{2}~=~-dt^{2}~+~cos^{2}(t)d\chi^{2}~+~cos^{2}(t)sinh^{2}(\chi)d\theta^{2}~+~cos^{2}(t)sinh^{2}(\chi)sin^{2}(\theta)d\phi^{2}
\]
for $d\phi~=~0$ and $cos(t)~=~cos(at)$ with $a~=~0$ reduces to
a three dimensional space with the metric $ds^{2}~=~-dt^{2}~+~d\chi^{2}~+~sinh^{2}(\chi)d\theta^{2}$.
We set the $T$ and $X$ coordinates so that $\chi~=~tanh^{-1}(T/X)$
and $T~=~x~sinh(\chi)$, $X~=~x~cosh(\chi)$, for $x~=~\sqrt{X^{2}~-~T^{2}}$.
This gives the Poincaré half-plane on a times slice with the metric

\[
ds^{2}~=~\frac{dx^{2}~+~dy^{2}}{y^{2}},~y~>~0
\]
It is easily shown that $\Gamma_{xy}^{x}~=$$\Gamma_{yy}^{y}~=~-1/y$,
and $\Gamma_{xx}^{y}~=~1/y$. The nonzero curvature components in
these coordinates are

\[
R_{xyx}^{x}~=~-\frac{1}{y^{2}},~R_{xxx}^{x}~=~R_{xyx}^{y}~=~-\frac{1}{y^{2}}
\]
With the Ricci curvatures $R_{11}~=$$R_{22}~=~-1/y^{2}$, Ricci scalar
curvature $R~=~-2$ and Gaussian curvature $K~=~-1$. This by way
of elementary illustration indicates the curvature is negative throughout
the space. The curvature approaches zero as $y~\rightarrow~\infty$and
diverges as $y~\rightarrow~0$. 

The geodesics of the Poincaré half plane are circles which perpendicularly
intersects at $y~=~$0 metric. The line element is then $ds~\simeq~dy/y$
with a small increment $s~\simeq~ln(y)$ . This diverges at zero,
where near this point we write the logarithm for $y~=~1~-~x$ so $ln(1~-~x)~=~-\sum_{n}x^{n}/n$.
For $x~=~\epsilon~-~1$, so that $x^{n}~\simeq~1~+~n\epsilon$ the
Taylor series is

\[
ln(1~-~x)~\simeq~-\sum_{n}\frac{1~+~n\epsilon}{n}~=~-\sum_{n}\left(\frac{1}{n}~+~\epsilon\right)
\]
This constructs the discrete form of the logarithmic divergence as
$\epsilon~\rightarrow~0$. Now substitute $n~\rightarrow~ne^{(n-1)\epsilon}$
with the implied limit $\epsilon~\rightarrow~0$ so that

\[
s~\simeq~ln(y)~=~-\sum_{n}\frac{e^{-(n-1)\epsilon}}{n}~=~-\int\sum_{n}e^{-(n-1)\epsilon}d\epsilon.
\]
This is a geometric series, 

\[
\sum_{m}e^{-m\epsilon}~=~e^{-\epsilon}(1~+~e^{-\epsilon}~+~e^{-2\epsilon}~+~\dots)~=~\frac{e^{-\epsilon}}{1~-~e^{-\epsilon}}
\]
The integral of this is $ln(1~-~e^{-\epsilon})~-~\epsilon$ or with
the logarithm Taylor series $\simeq~-2\epsilon$. Hence the line element
is regular.

We now turn our attention to curvature. The line element in a small
neighborhood of size $\epsilon$, a variation of the line element
is 

\[
s~=~s_{0}~+~\epsilon\frac{ds}{d\epsilon}~+~\epsilon^{2}\frac{1}{2}\frac{d^{2}s}{d\epsilon^{2}}~+~\dots
\]
It is clear that the second order term contains curvature information
from

\[
\frac{d^{2}s}{d\epsilon^{2}}~=~\nabla_{i}\nabla_{j}g_{kl}\frac{dx^{i}}{d\epsilon}\frac{dx^{j}}{d\epsilon}dx^{k}dx^{l}~=~\frac{1}{2}[\nabla_{i},\nabla_{j}]g_{kl}\frac{dx^{i}}{d\epsilon}\frac{dx^{j}}{d\epsilon}dx^{k}dx^{l}
\]
where the commutator is the components of a curvature two-form. The
essential information to describe the behavior of curvature near the
boundary is the second derivative of $ln(y)$ or equivalently

\[
\frac{d^{2}s}{d\epsilon^{2}}~=~-\frac{d}{d\epsilon}\sum_{n}e^{-(n-1)\epsilon}~=~-\frac{d}{d\epsilon}\frac{e^{-\epsilon}}{1~-~e^{-\epsilon}}
\]
The term differentiated is expanded in a Taylor series for small $\epsilon$
so $e^{-\epsilon}~~=~1~\text{\textendash}~\epsilon$ $+~\epsilon^{2}/2$
and this summation, with use of the binomial theorem and eliminating
powers $O(\epsilon^{3})$ and higher, is

\[
\sum_{m}e^{-m\epsilon}~=~-\frac{1}{\epsilon}(1~\text{\textendash}~\epsilon/2~-~\epsilon^{2}/12).
\]
Now take the derivative with elementary calculus rules change the
sign as above and we have 

\[
\frac{d^{2}s}{d\epsilon^{2}}~=~-\frac{1}{\epsilon^{2}}~+~\frac{1}{12}.
\]
As $\epsilon~\rightarrow~0$ the first term blows up. This UV divergence
can be absorbed into the definition of the momentum and regularized
away. The remaining term is the value of the curvature on the boundary,
which is finite and positive.

The physical interpretation is the quantum vacuum energy. The vacuum
energy of quantum fields in spacetime is $E~=~D\omega_{0}\sum{}_{n}n/2$,
in a box normalization, and with $D$ the dimension of the harmonic
oscillators. The sum $1~+~2~+~3~+~\dots$ $+~n~+~\dots$ is equal
to 

\[
\sum_{n}n~=~\frac{d}{d\epsilon}\frac{e^{-\epsilon}}{1~-~e^{-\epsilon}}
\]
which means the vacuum energy near the boundary is equal to

\[
E~=~D\omega_{0}\left(\frac{1}{2\epsilon^{2}}~-~\frac{1}{24}\right)
\]
The regularized vacuum energy requires that $D~=~24$. This is the
spatial dimension of the oscillator, which is in a light cone (light
front) frame, and so the number of spatial dimension is $25$ and
the spacetime dimension is $26$. The vacuum energy contributes a
curvature term that is $\Lambda~\sim~\omega_{0}$, which is positive.
The sign is changed by the negative curvature of the $AdS_{4}$, which
in the evaluation of the vacuum state on $\partial AdS_{4}$ changes
the sign so $\Lambda~\propto~-12\omega_{0}\sum_{n}n$.

The boundary of the anti-de Sitter spacetime is more general than
this elementary case with the Poincaré disk. The boundary of the $AdS_{5}$,
$\partial AdS_{5}~=~E_{c}^{4}$, is a conformally flat spacetime {[}7{]}.
In line with the even dimensional construction we may for instance
consider $AdS_{6}$, $\partial AdS_{6}~=~E_{c}^{5},$with a compactification
of one dimension of $E_{c}^{5}$ into $E_{c}^{4}\times S^{1}$. The
conformal transformation $g_{\mu\nu}~\rightarrow~\Omega^{2}g_{\mu\nu}$
the flat spacetime element $ds^{2}~=~du^{2}~-~\sum_{i}dx^{i}dx^{i}$
is a time dependent transformation for $du/dt~=~\Omega^{-1}$ and
a de Sitter spacetime for $\Omega^{2}~=~e^{\sqrt{\Lambda/3}~t}$.
This is approximately the spacetime for our physical universe. It
is then argued that the negative cosmological constant in the $AdS_{5}$
spacetime may manifest itself as a positive cosmological constant
on the boundary.

\section{Geometric quantization of brane-world?}

The $AdS_{n+1}$ group of isometries $O(n,2)$ contains a Möbius subgroup,
or modular transformations, so this discrete group does not necessarily
act effectively on $AdS_{n+1}$. This means that the discrete group
$\Gamma$ is not necessarily convergent on the boundary space $M_{n}$.
Such a convergence means there exists a sequence $g_{i}~\in~\Gamma$
which admits a north-south dynamics of poles $p^{\pm}$ on a sphere,
which in the hyperbolic case defines the past and future portions
of a light cone {[}8{]}. The limit set of a discrete group is a closed
$\Gamma$-invariant subset that defines a $\Lambda_{\Gamma}~\subset~M_{n}$
so the complement $\Omega_{\Gamma}$ acts properly on $M_{n}$. This
$\Gamma$-invariant closed subset of $\Lambda_{\Gamma}~\subset~L_{n}$
is the space of lightlike geodesic in $M_{n}$. The action of $\Gamma$
on $\Omega_{\Gamma}\cup AdS_{n+1}$ is contained in $M_{n}$. The
open set $\Lambda_{\Gamma}$ is the maximal set that the $\Gamma$
acts properly on $\Omega_{\Gamma}\cup AdS_{n+1}$. The other is the
discrete group $\Gamma$ is Zariski dense in $O(n,~2)$.

The lightlike geodesics in $M_{n}$ are copies of $\mathbb{R}P^{1}$,
which at a given point $p$ define a set that is the light cone $C(p)$
{[}8{]}. The point $p$ is the projective action of $\pi(v)$ for
$v$ a vector in a local patch $\mathbb{R}^{n,2}$ and so $C(p)$
is then $\pi(P\cap C^{n,2})$, for $P$ normal to $v$, and $C^{n,2}$
the region on $\mathbb{R}^{n,2}$ where the interval vanishes. The
space of lightlike geodesics is a set of invariants and then due to
a stabilizer on $O(n,2)$, so the space of lightlike curves $L_{n}$
is identified with the quotient $O(n,2)/P$, where $P$ is a subgroup
defined the quotient between a subgroup with a Zariski topology, or
a Borel subgroup, and the main group $G~=~O(n,~2)$. This quotient
$G/P$ is a projective algebraic variety, or flag manifold and $P$
is a parabolic subgroup. The natural embedding of a group $H~\rightarrow~G$
composed with the projective variety $G~\rightarrow~G/P$ is an isomorphism
between the $H$ and $G/P$. This is then a semi-direct product $G~=~P~\rtimes~H$.
For the $G$ any $GL(n)$ the parabolic group is a subgroup of upper
triangular matrices, called Borel groups {[}9{]}.

The connection between the symplectic group symplectic group $Sp(2N_{f})~\subset$
$SU(2N_{f})~\sim~SO(2,~2)$ and the parabolic group of upper triangular
matrices is a geometric quantization {[}10{]}. The symplectic manifold
$(M,~\Omega)$, where $\Omega$ is the skew symmetric matrix defines
a prequantization as a representation of elements of the Poisson algebra
$f~\in~C^{\infty}(M)$ as sections of a Kahler line bundle $L$, with
$\pi:L~\rightarrow~M$. The prequantization line bundle contains the
one form $\omega~=~df~+~2\pi i\alpha$, for $\alpha$ on the line
bundle, such that the curvature $R~=~D\wedge D$, for $D~=~d~+~\omega$
under $\pi:R~=~i\Omega$. Let $T~=~T(M)$ be the tangent bundle to
$M$ for elements $u,~v~\in~T$. The Poisson bracket $\{u,~v\}~\in~\Gamma(T)$
exists on sections of $T$, and the quantum algebra $Q_{M}$ of $M$
is the operators formed from functions $f$ such that their Hamiltonian
vector fields $x_{f}$ $x^{a}~=~\Omega^{ab}\partial_{b}H$, and $[x_{f},~T]~\subset~T$.
$Q_{M}$ forms a pre-Hilbert space of half-forms, tensor density fields
with weight $s~=~1/2$, with $f_{op}~=~f~+~i\hbar^{1/2}{\cal L}_{T}x_{f}$,
with $\alpha~=~i\hbar^{1/2}{\cal L}_{T}x_{f}$. This is a form of
the $\star$-product that is an extension of the function on a symplectic
manifold into a quantum algebra. 

This gives two routes to quantization. The first is with a geometric
quantization approach with the enveloping algebra on $U,~V$ and $Sp(2N_{f})$
for a D-brane, the other is with the conformal completion of the $AdS_{n}$.
In the latter case the parabolic subgroup of Borel groups or Heisenberg
groups. The parabolic group defines light cones, which are an invariant
of spacetime. The invariance of spacetime is proper time, where in
this construction the proper time is zero. In the geometric quantization
approach the coordinates of phase space are employed, and Hamiltonian
vector fields $u^{a}~=~d\gamma^{a}/dt$ are defined according to a
coordinate definition of time. Quantum fields in spacetime are defined
according to local operators that commute on a spatial surface of
simultaneity. Hence QFT is defined according to coordinate time. In
one case the quantization $\star$-product is constructed according
to light cones, or proper intervals, which is more commensurate with
the structure of general relativity. In the braney approach the quantization
is tied to coordinate geometry and is in line with quantum field theory.
These rely entirely on different definitions of time. The open question
is then how are these related, whether they ultimately give identical
results, or whether these two schemes are aspects of a more general
quantum gravity scheme.

\section{Concluding statements}

The braney dynamics with strings is a form of QCD, where the endpoints
of a string are {}``quarks'' with a color identified with the brane
it is anchored to. The QCD dynamics of the brane with $SU(2,~2)$
symmetry, or $SO(4,~2)$, is governed by the noncommutative geometry
of the $\star$-product. The development of a brane from the bifurcation
of a string is a form of infinite momentum boost which increases the
number of degrees of freedom on the brane. As the number of modes
increases the brane becomes a classical-like object. The $SU(2,~2)~\sim~SO(4,~2)$
is the isometry group for the $AdS_{4}$, and acts as a QCD-like gauge
field. The decomposition of $SU(2,~2)$ into $Sp(4)~\sim~Sp(2,~2)$
form the symplectic basis for the noncommutative geometry of the brane,
or AdS-brane. 

The observable universe is likely connected to AdS spacetime. Hartle,
Hawking and Hertog argue the physical universe may indeed by anti-de
Sitter. As with the Poincaré half plane, or the Poincaré disk, the
geodesics are great arcs which leave the boundary with enormous curvature
and high energy, traverse the space and return to the boundary. An
observer in an anti-de Sitter spacetime would observe distant objects
to be highly blue shifted. Any object observed at a great distance
would emit radiation which is blue shifted towards the observer. It
is for this reason the anti-de Sitter spacetime was considered in
quantum gravity, for this property makes it the perfect box to hold
a black hole. The boundary has a repellant gravitational influence.
The argument is made for why the observable universe is a conformally
flat spacetime on the boundary of the $AdS_{n}$spacetime. 

The noncommutative geometry of the brane, or geometric quantization,
then shares some relationship with the conformal completion of the
$AdS_{n}$ spacetime and the Borel group upper triangular matrix form
of the Heisenberg group. The two approaches then share some relationship
which is as yet not clear. It could be the two forms of quantization
are not equivalent and then must embed in some more general form of
quantum gravity.

\section{References}

{[}1{]} L. B. Crowell, {}``Tricritical quantum point and inflationary
cosmology,'' This essay received an honorable mention in the 2012
Essay Competition of the Gravity Research Foundation. submitted to
IJMPD arXiv:1205.4710v1 {[}gr-qc{]}\vskip.03in\noindent{[}2{]} A. Pich, {}``Quantum Chromodynamics,'' arXiv:hep-ph/9505231v1\vskip.03in\noindent{[}3{]} T. Banks, W. Fischler, S. H. Shenker, L. Susskind, {}``M
Theory As A Matrix Model: A Conjecture,'' ${\it Phys.~Rev.}$, ${\bf 55}$
5112-5128, 1997, 5112-5128 arXiv:hep-th/9610043v3\vskip.03in\noindent{[}4{]} V. G. Kac, {}``Highest weight representations of infinite dimensional Lie algebras,'' Proc. Internat. Congress Mathematicians, Helsinki, 1978\vskip.03in\noindent{[}5{]} L. Susskind, \char`\"{}Strings, Black Holes and Lorentz Contraction,\char`\"{}
${\it Phys.~Rev.}$, ${\bf D49}$, 6606-6611 (1994) arXiv:hep-th/9308139v1\vskip.03in\noindent{[}6{]} J. B. Hartle, S. W. Hawking, T. Hertog, {}``Accelerated Expansion from Negative$\Lambda$,'' http://arxiv.org/abs/1205.3807v2\vskip.03in\noindent{[}7{]} J. M. Maldacena, {}``The Large N Limit of Superconformal Field Theories and Supergravity,\char`\"{} ${\it Adv.~Theor.~Math.~Phys.}$, ${\bf 2}$, 231-252 (1998). \vskip.03in\noindent{[}8{]} C. Frances, {}``Lorentzian Kleinian Groups,'' ${\it Comment.~Math.~Helv.}$${\bf 80}$,4, 883-910 (2005) . \vskip.03in\noindent{[}9{]} A. Borel, {}``Groupes linéaires algébriques,'' ${\it Ann.~of~Math.}$,${\bf 6}$, 4, (1956) \vskip.03in\noindent{[}10{]} N. M. J. Woodhouse, ${\it Geometric~Quantization}$,'' Clarendon Press, 1991.
\end{document}